\documentclass[conference]{IEEEtran}
\usepackage{times,url,comment,cite}
\usepackage{amsmath,amssymb,amsfonts,bbm}
\usepackage{graphicx,slashbox}
\usepackage{epsfig,epsf}
\usepackage{color}
\usepackage{subfigure}

\makeatletter
\def\ps@headings{%
\def\@oddhead{\mbox{}\scriptsize\rightmark \hfil \thepage}%
\def\@evenhead{\scriptsize\thepage \hfil \leftmark\mbox{}}%
\def\@oddfoot{}%
\def\@evenfoot{}}
\makeatother

\usepackage{cite}

\ifCLASSINFOpdf
\else
\fi
\usepackage{url}

\definecolor{marco}{rgb}{0,0.6,0}

\begin{document}

\newtheorem{theorem}{Theorem}[section]
\newtheorem{lemma}{Lemma}[section]

\title{A Holistic Approach to Information Distribution\\ in Ad Hoc Networks}

\author{\IEEEauthorblockN{Claudio Casetti, Carla-Fabiana Chiasserini, 
Marco Fiore}
\IEEEauthorblockA{Politecnico di Torino, Italy
}
\and
\IEEEauthorblockN{Chi-Anh La, Pietro Michiardi}
\IEEEauthorblockA{EURECOM, France}
}


\maketitle

\begin{abstract}
We investigate the problem of spreading information contents 
in a wireless ad hoc network with mechanisms embracing the peer-to-peer
paradigm. In our vision, information dissemination should satisfy 
the following requirements: (i) it conforms to a predefined distribution and 
(ii) it is evenly and fairly carried by all nodes in their turn. 
In this paper, we observe the dissemination effects 
when the information moves across nodes according to two well-known
mobility models, namely random walk and random direction. 
Our approach is fully distributed and comes at a very low cost in terms of protocol overhead;
in addition, simulation  results show that
the proposed solution can achieve the aforementioned goals under different network scenarios,
provided that a sufficient number of information replicas are injected into the
network. This observation calls for a further step: in the realistic case where the user content demand 
varies over time, we need a content replication/drop 
strategy to adapt the number of information replicas to
the changes in the information query rate. We therefore devise a distributed, lightweight
scheme that performs efficiently in a variety of scenarios. 

\end{abstract}


\section{Introduction}

It is commonly acknowledged that user devices are rapidly becoming 
tantamount to a communication hub, sporting
arrays of GPS navigators, web browsers, videogame consoles and 
screens flashing the latest news or local sightseeing information. 
In this context, most pieces of information are likely to be of general use,
and therefore a sensible dissemination and caching policy would be 
desirable. 

In this work, we focus on such an environment: few and far between 
access points, or gateway nodes, in a highly-populated network area
where user devices are equipped with a data cache and 
communicate through the ad hoc networking paradigm.
Users create a cooperative environment  
where information is exchanged among nodes in a peer-to-peer fashion.
In particular, they form a pure peer-to-peer system, whose nodes
may simultaneously act as both ``clients" and ``servers" 
to the other nodes in the network.
Also, we envision a system that achieves a desired  distribution over the network 
of the information that users
may be interested in. 
By information distribution, we mean
the distribution according to which the information copies 
should be cached in the network.
The nodes storing an information copy will act as {\em providers} for
this content. 

Traditional approaches to information caching in communication 
networks~\cite{Baev01,Prabh05,Nuggehalli06,Tang08}
are based on the solution of linear programming problems,
which often require global knowledge of the network condition, or lead
to quite complex solutions that involve significant communication overhead.  
Unlike previous approaches, our solution 
is  fully distributed and it comes at a very low cost in terms of
communication overhead. 
Our goal is to achieve the desired content distribution
by properly letting the information move across the network.

More specifically, while developing our solution, 
we identify a number of 
issues that need to be addressed.
\begin{itemize}
\item {\it Achieving the desired distribution of the information:}
regardless of how the information is distributed at the outset, the system
should be able to identify 
where 
the information should be 
stored in the network.

\item {\it Fair distribution of information burden:} 
as mentioned above, a node storing
the information acts as provider for that information; of course, this role may
exact a high toll from nodal resources in terms of bandwidth or power
consumption; it is therefore advisable that the role of content provider be
handed over to neighboring nodes quite frequently, without altering the
information distribution.

\item 
{\it Dynamic adaptation to time-varying content demands:}
given an initial number of information replicas, which implies a fair load distribution among
the provider nodes, the system should be able to adapt to the varying information query rate,
by increasing or decreasing the number of information copies in the network as needed.


\end{itemize}

We deal with the above issues
and develop a solution that features the following advantages:
\begin{itemize}
\item it is fully distributed;
\item it is content-transparent, i.e., 
it does not require knowledge of the contents stored by
the neighboring users;
\item it works with minimum overhead.

\end{itemize}

In particular, motivated by the need of a balanced load distribution 
among the provider nodes and of an equal quality of service provisioning to the users,
we target a uniform distribution of  
contents, either  over 
the network spatial area or over the network nodes.
With this aim in mind, in Section~\ref{sec:info-distribution}
we investigate the applicability of two 
well-known mobility models, namely the random walk and the 
random direction model, to disseminate the information across the
network.  Both strategies, using the experimental setup in 
Section~\ref{sec:set-up}, are proven to yield a  distribution 
of the information copies that is close to the
target distribution, regardless of the considered network scenario 
(Section~\ref{sec:results}).  
Also, the obtained results show that the level of fairness in distributing 
the burden among provider nodes
depends on the number of information copies stored in the network.
Thus, when the user query rate for a content varies 
over time, we address the problem of
how to let the number of content replicas adapt to the changing content demand.
To address this issue, in Section~\ref{sec:replication-drop} we devise
a content replication/drop algorithm that 
controls the number of information copies in the network, taking into account 
the time-varying behavior of the contents popularity level.
The performance of our scheme is presented in Section~\ref{sec:results_replication}.


\section{Related work}
\label{sec:rel-work}

Our study is related to the problem of optimal cache placement 
in wireless networks. Several works have
addressed this issue by exploiting its similarity 
 to the facility location and the $k$-median problems. Both
these problems are NP-hard and a number of constant-factor 
approximation algorithms have been proposed for each of them 
\cite{Jain01,Baev01,Nuggehalli06}; these algorithms however are not amenable
to an efficient distributed implementation.

Distributed algorithms for allocation of information replicas
are proposed, among others, in \cite{Hara01,Tang08,Cao04,Holliday03}.
These solutions typically involve significant communication overhead,
especially when applied to mobile environments, and focus on 
minimizing the information access cost or the query delay.  In our work,
instead, we consider a cooperative environment and aim at a uniform 
distribution of the information copies, while evenly distributing 
the load among the nodes acting as providers.

Relevant to our study is also the work in \cite{Aazami04},
which computes the (near) optimal number of replicas of video clips in wireless
networks, based
on the bandwidth required for clip display and their access statistics.
However, the strategy proposed in \cite{Aazami04} requires 
a centralized implementation and applies only to strip or grid topologies.
In the context of sensor networks, the study in \cite{Kumar04}
analytically derives the minimum number of sensors
that ensure full coverage of an area of interest, 
under the assumption of a uniform sensor deployment.

Again in the context of sensor networks, approaches based on
active queries following a trajectory through the network,
or agents propagating information on local events have been 
proposed, respectively, in \cite{Helmy05} and \cite{Estrin02}.
Note that both these works focus on the forwarding of these
messages through the network, while our scope is to make 
the desired information available by letting it move through
nodes caches. 


\section{Achieving the desired information distribution}
\label{sec:info-distribution}

We start by addressing the problem of where the information copies should 
be cached in the network so as to obtain the desired content distribution.

We consider a tagged\footnote{I.e., 
we assume information to be uniquely identifiable.}
information and we take the desired
distribution to be uniform, over either the network area ({\em spatial uniformity})
or the network nodes ({\em nodal uniformity}).  
Spatial uniformity is motivated by the need
of providing the same quality of service (e.g., 
probability of finding the content and information delivery delay)
to all network users, while nodal uniformity fosters 
load balancing among the nodes acting as content providers.

To achieve the target distribution, we let the information move across
nodes according to two well-known mobility models, namely the random walk~\cite{einstein26}
and the random direction~\cite{Nain05}   models, which are
often used to represent the movement of user nodes in ad hoc networks. 
In our context, a mobile entity is not a network node but, rather, a copy of the tagged information which
``hops'' from a user node that just stopped being a provider for that information
onto another node which will become the new content provider.
We apply the two mobility models and develop the dissemination strategies 
detailed below. 

\begin{itemize}
\item {\bf The random walk dissemination (RWD) strategy:}
we consider the simplest random walk possible, in which each mobile entity,
i.e., an information copy, roams the network by moving from one node to a
one-hop neighbor selected with equal probability.
Each node caches the information 
before handing it over to the next hop in the information copy visit pattern.
This approach requires trivial node operations and introduces minimal
overhead, thus representing a lower-bound benchmark for more advanced
information mobility models.

\item{\bf The random direction dissemination (RDD) strategy:} it implies that each
mobile entity alternates periods of movement (move phase) to periods during
which it pauses (pause phase). In our context, the pause phase corresponds to
the time period during which the information copy is stored at a provider node.
The move phase starts at the time instant when the current information
provider hands over the content to one of its one-hop neighbors, 
and it ends when the new provider is reached by the information copy.
At the beginning of a move phase, the current provider independently selects
the direction and the distance\footnote{Note that randomly selecting a travel distance 
is equivalent to randomly selecting speed and travel time.} 
for the movement of the information copy, thus
identifying a target location whose position is included in the content
messages.
An application-driven routing allows the information to be moved towards
the target location, with each forwarder selecting as a next hop the neighbor
that best fits the ideal trajectory designed by the original provider.
The neighbor selection process is performed in a reactive manner, as it
involves an exchange of advertisement (by the forwarder) and reply (by
candidate next hop neighbors) messages at each movement hop.
When a node has no neighbors closer than itself to the target position,
it elects itself as the new provider, and the pause phase starts again.
Some remarks are in order. First, this scheme requires that user nodes
be capable to estimate their position (i.e., through GPS), a fair assumption in
most practical scenarios.
Second, the information moves across user nodes, thus it may be transmitted
along a direction that just approximates the planned trajectory, or it may be
stored at a node that is nearby (but not exactly at) the selected geographical
destination.
Third, geographical areas devoid of nodes that can support the information
movement may be encountered during move phases: in that case, the current
forwarder assumes a {\it boundary} has been hit, and applies a
reflection to the movement angle.
\end{itemize}

As already mentioned, using the RWD and RDD strategies translates 
into a fully-distributed, low-overhead solution. Furthermore,
in the literature there are results showing that the random walk and the random direction mobility models
can lead to a uniform distribution of the mobile entities.
Indeed, if the network topology can be represented as 
an undirected, connected, non-bipartite graph, then the distribution of 
entities moving according to the random walk model converges to a unique
stationary distribution regardless of the initial distribution, and 
this stationary distribution is uniform in the case of regular graphs%
\footnote{A graph is regular if each of its vertices has the same 
number of neighbors.} 
\cite{CRW07}. 
As for the random direction model, in \cite{Nain05} it has been shown that, if at time $t=0$ the position and the
orientation of mobile entities are independent and uniform over
a finite square area, they remain uniformly distributed over the area for all time instants $t>0$,
provided that the entities move independently of each other.

However, in practice: 
(i) wireless ad hoc networks often have an irregular structure, even
changing over time if the users are mobile, thus the results in \cite{CRW07}
for the random walk model do not directly apply; (ii)  in the case of the RDD strategy,  
the information only approximately reaches its geographical destination,
as noted above; (iii) we need to address both spatial and nodal uniformity. 
Therefore, in the following,  we investigate the actual  distribution of
the information that is obtained through our approach and how far it is from the
target distribution. 

\section{Experimental set-up and methodology}
\label{sec:set-up}

Before delving into the evaluation of the techniques we propose
for information dissemination, we first describe our 
experimental settings.

We use the \emph{ns}-2 network simulator, where all nodes are equipped with 
standard 802.11b interfaces, with 11~Mbps fixed data transmission rate. We 
enhanced the simulator with the random walk and the random direction
dissemination algorithms, described in 
Section~\ref{sec:info-distribution}. We also implemented a simple application that 
allows mobile nodes to query provider nodes. Any node wishing to access a copy 
of the information maintained in the network executes a limited-scope flooding 
of a query message. Queries can traverse a maximum number of hops, $h_{max}=5$, 
before being discarded. We improve the query propagation process by adopting the 
PGB technique~\cite{naumov06} to select forwarding nodes that relay 
queries to their destinations. Sequence numbers are used to detect and discard 
duplicate queries and avoid the \emph{broadcast storm} 
phenomenon~\cite{storm99}. Upon reception of a query, a provider replies with a probability 
that is inversely proportional to the number of hops traversed by the query 
message. Note that it is out of the scope of this paper to design an optimized 
application to access information: nevertheless, our simple design is sufficient 
to evaluate the information dissemination algorithms discussed in 
Section~\ref{sec:info-distribution}. 

In the following, we define the simulation settings we analyze in this work. 
Note that all results presented in the remainder of this paper are averaged out 
over 10 simulation runs, each with a randomized selection of initial information 
providers. Simulation time is set to 10,000 seconds, unless specified otherwise. Moreover, we assume a 
network composed of $N = 2000$ nodes that are spatially distributed on a square 
area $\mathcal{A}$ of $500$ m side. Each node has a transmission range of 20~m.
When employing the RDD scheme, providers characterize the information move phase
by randomly choosing angles that are uniformly distributed in $[0,2\pi]$, and 
exponentially distributed distances, with mean 100~m.
We study both \emph{static} and \emph{mobile} cases, as will be detailed below.

\subsection{Static node spatial distribution}

\begin{figure*}[!th]
  \begin{tabular}{ccc}

    \begin{minipage}[t]{0.3\textwidth}
      \begin{center}    
        \subfigure[Uniform]{\includegraphics[scale=0.4]{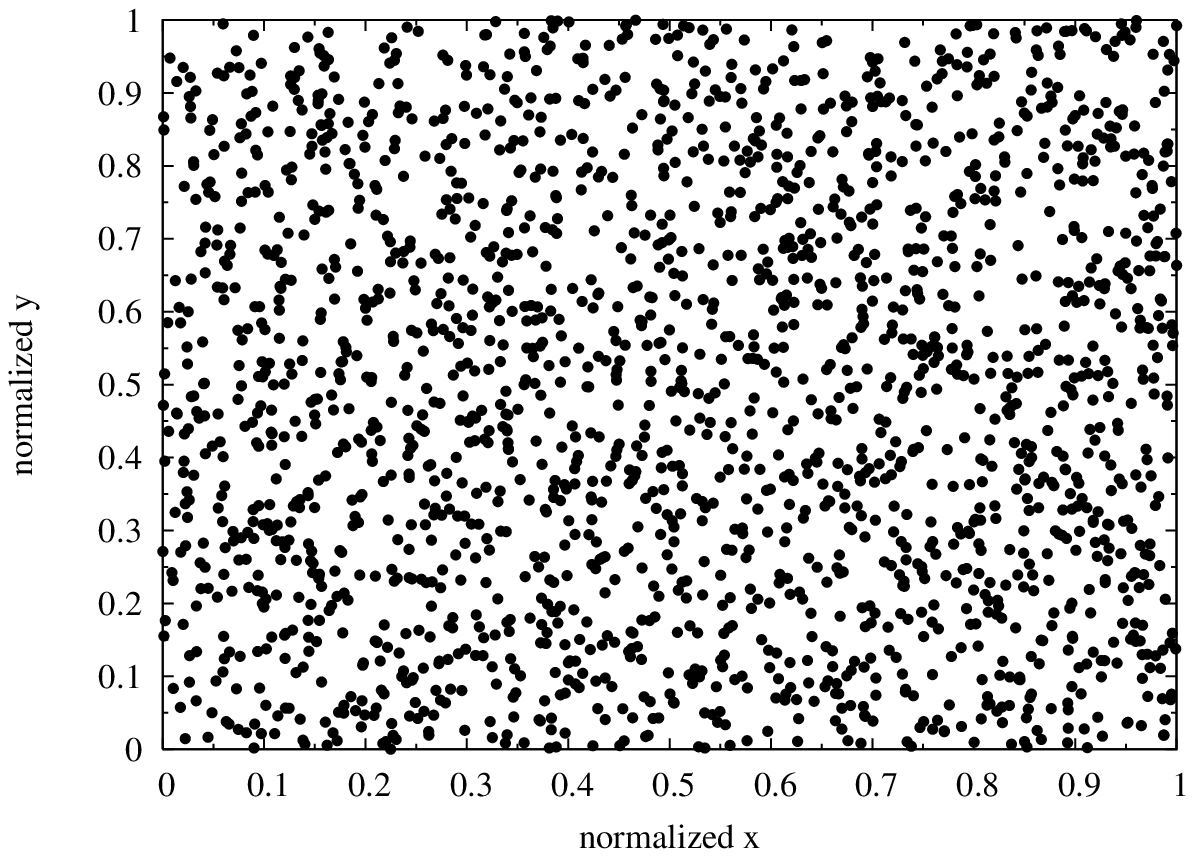}}
      \end{center}
    \end{minipage}

    &

    \begin{minipage}[t]{0.3\textwidth}
      \begin{center}  
        \subfigure[Stationary]{\includegraphics[scale=0.4]{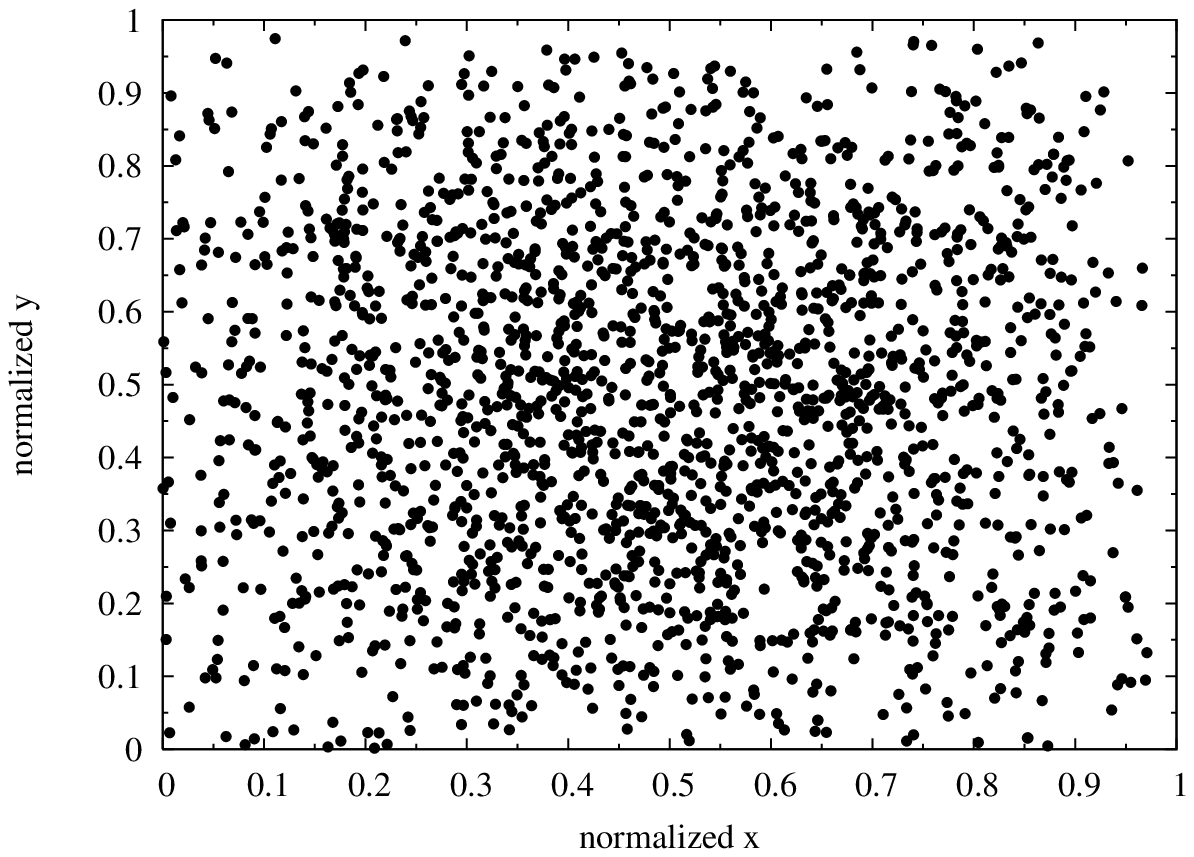}}
      \end{center}
    \end{minipage}

    &

    \begin{minipage}[t]{0.3\textwidth}
      \begin{center}  
        \subfigure[Clustered]{\includegraphics[scale=0.4]{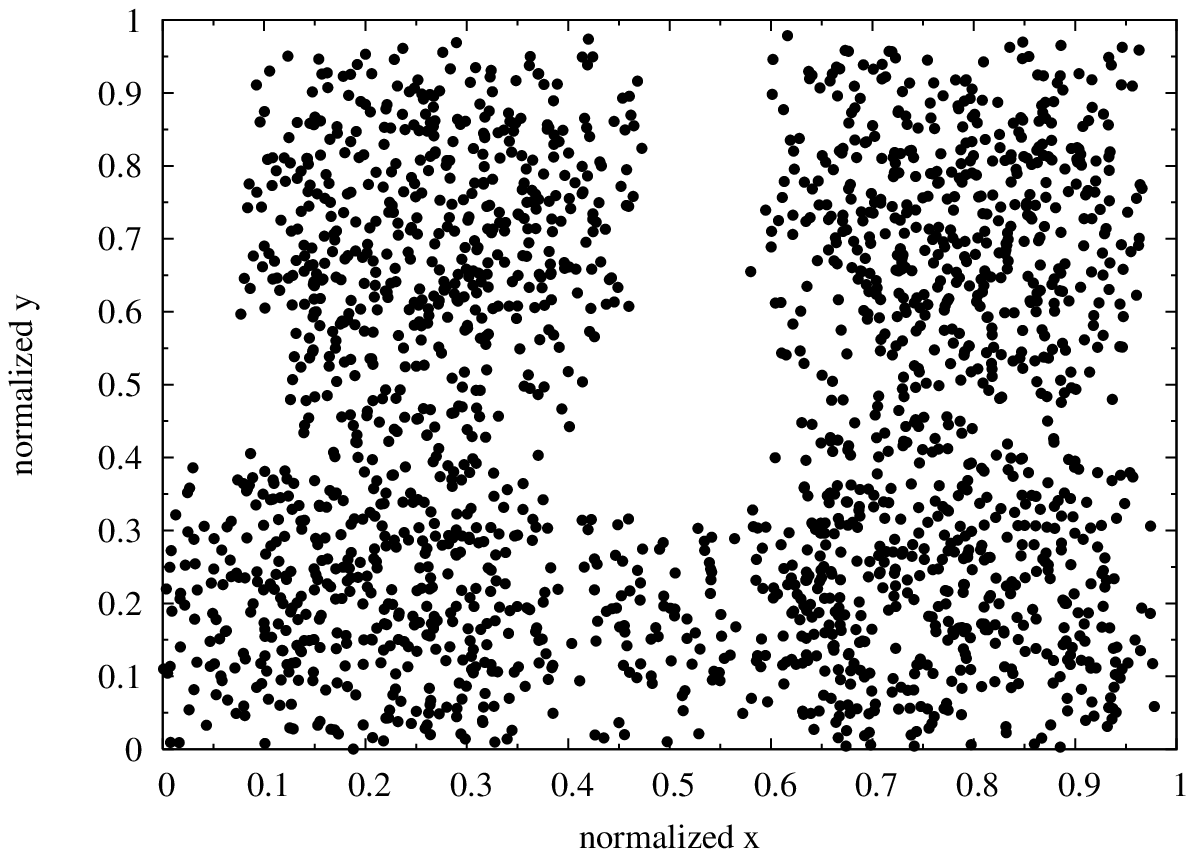}}
      \end{center}
    \end{minipage}
   \end{tabular}
  
\caption{Snapshots of node deployment used in our experiments.}    
  \label{fig:topo}
 \end{figure*}

We define the following static node deployments, samples of which 
are depicted in Fig.~\ref{fig:topo}:

\begin{itemize}
\item \emph{Uniform distribution}: nodes are uniformly placed
on $\mathcal{A}$;

\item \emph{Stationary distribution}: 
we consider a deployment, where, as discussed in \cite{leboudec2005}, nodes are more often 
located towards the center of the network area; 

\item \emph{Clustered distribution}: we assume nodes to be deployed in four equally 
sized clusters. Each cluster corresponds to a ``point of interest'' 
around which nodes are located. 
Nodes are also placed in-between clusters so as to ensure network connectivity. 
In practice, we implement the random trip model as defined in 
\cite{leboudec2006} and take a snapshot of the network topology as our initial 
node distribution. 
\end{itemize}

\subsection{Node mobility}
The impact of node mobility on the dissemination mechanisms we
designed is analyzed for the following mobility models:
\begin{itemize}

\item \emph{Stationary Random Waypoint}: as discussed in \cite{leboudec2005}, we 
enhance the basic random waypoint model so as to reduce the transient phase of 
our analysis. In this setting, the initial node distribution is the same as the 
stationary distribution described for the static case; each node selects a 
random destination in $\mathcal{A}$ and moves towards it at a constant speed 
selected uniformly at random from a distribution with a mean of 3 m/s. The pause 
time is set to 10 s.

\item \emph{Random trip}: following the definition in \cite{leboudec2006}, nodes 
revolve around four ``points of interest''. The initial node deployment conforms to 
the clustered distribution defined for the static case. The stationary random 
waypoint model defined above guides node movements inside a cluster. 
Inter-cluster mobility is allowed with probability 0.3.


\end{itemize}

\subsection{Parameter space}
\label{sec:paramters}

We now define the parameters used in our evaluation, accounting for the initial 
distribution and number of information providers in the network, as well as the 
query behavior of mobile nodes.

\begin{itemize}
\item \emph{Number of information providers}: at the beginning of
each simulation run, a predefined number of providers is 
randomly chosen among all nodes in the network. 
$\mathcal{C}(t)$ is the number of providers available at time $t$; 
we choose $\mathcal{C}(0) \in \{ 20, 50, 100, 200, 400\}$. 

\item \emph{Information caching time}: when taking up the role of 
information provider, a node $i$ keeps a local copy for a time $\tau_i$. 
In this work, we assume $\tau_i=\tau~ \forall i$ with $i=1,\ldots,N$. In the
following we present results for $\tau~\in~\{ 10, 100\}$ seconds.

\item \emph{Information demand}: we assume nodes to issue queries to information 
providers using the simple application defined above. Without loss of 
generality, we focus on one information content (of size 1KB) that is made 
available in the network. Users' demand for the available information 
is modeled through a query rate which we assume to be common to all users, 
$\lambda_i = \lambda = 0.0025 \text{ req/s } \forall i$ with $i=1,\ldots,N$. 
The aggregate query rate $\Lambda$ over all nodes depends on the 
number of information providers currently active in the 
network\footnote{Indeed, providers do not issue requests to access the
content}, i.e.,  $\Lambda(t)=(N-\mathcal{C}(t)) \lambda$. 
While our analysis mainly focuses on a constant user demand, 
we also introduce a more realistic scenario in which $\lambda$ 
varies over time. 

\end{itemize}

\subsection{Evaluation metrics}

To understand to which extent the information distribution achieved by our 
dissemination techniques resembles the desired content diffusion, we employ the 
well-known $\chi^2$ goodness-of-fit test on the inter-distance between 
information copies. As a matter of fact, we can compare the measured 
inter-distance distribution against the theoretical distribution of the distance 
between two points, whose position is a random variable following the target 
distribution. Using inter-distances instead of actual coordinates allows
us to handle a much larger number of samples (e.g., $C(t)(C(t)-1)$ instead of
just $C(t)$ samples) thus making the computation of the 
$\chi^2$ index more accurate.
As discussed before, we consider the following two reference distributions:
\begin{itemize}

\item {\em Spatial uniformity:}
since we consider a square area where nodes are deployed and 
we seek a uniform dissemination of content over the network area, the target distribution is the 
solution to the bidimensional case of the hypercube line picking 
problem~\cite{trott04}, which is known to be:
\[
q(x)=
\left\{
\begin{array}{l}
2x\left(x^2-4x+\pi\right) \\
\hspace{40mm} \text{if } 0 \leq x < 1, \\
2x\left[4\gamma-\left(x^2+2-\pi\right)-4\tan^{-1}\gamma\right] \\
\hspace{40mm} \text{if } 1 \leq x < \sqrt{2},
\end{array}
\right.
\]
with $\gamma = \sqrt{x^2-1}$.


\item {\em Nodal uniformity:} in order to test the 
uniformity over the network nodes, we take as a reference distribution the 
empiric distribution of node inter-distances measured in simulation.

\end{itemize}


\begin{figure*}[htbp]
  \begin{tabular}{cc}

    \begin{minipage}[t]{0.5\textwidth}
      \begin{center}    
        \subfigure[Static scenario]{\includegraphics[scale=0.3]{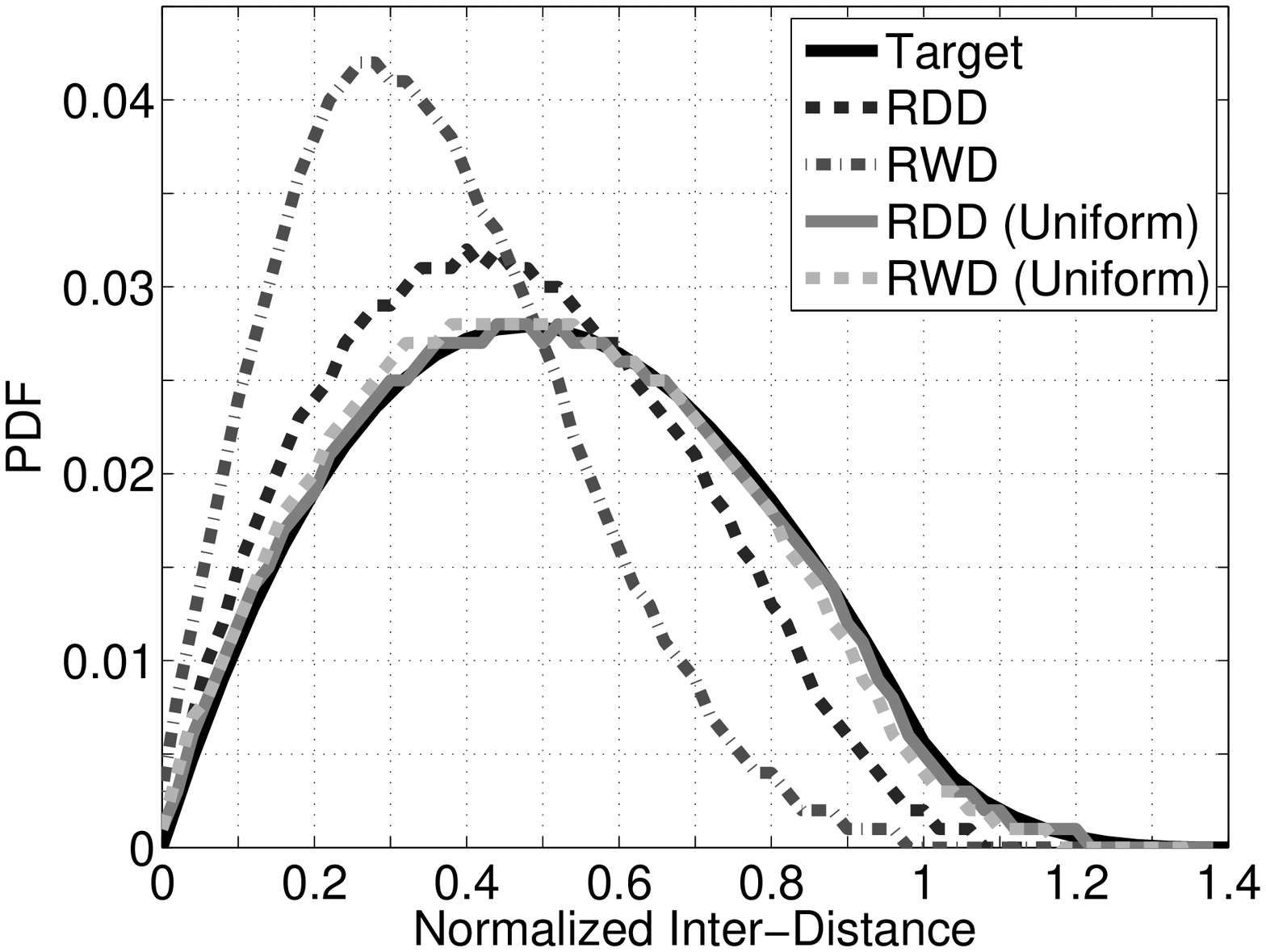}}
      \end{center}
    \end{minipage}

    &

    \begin{minipage}[t]{0.5\textwidth}
      \begin{center}  
        \subfigure[Mobile scenario]{\includegraphics[scale=0.3]{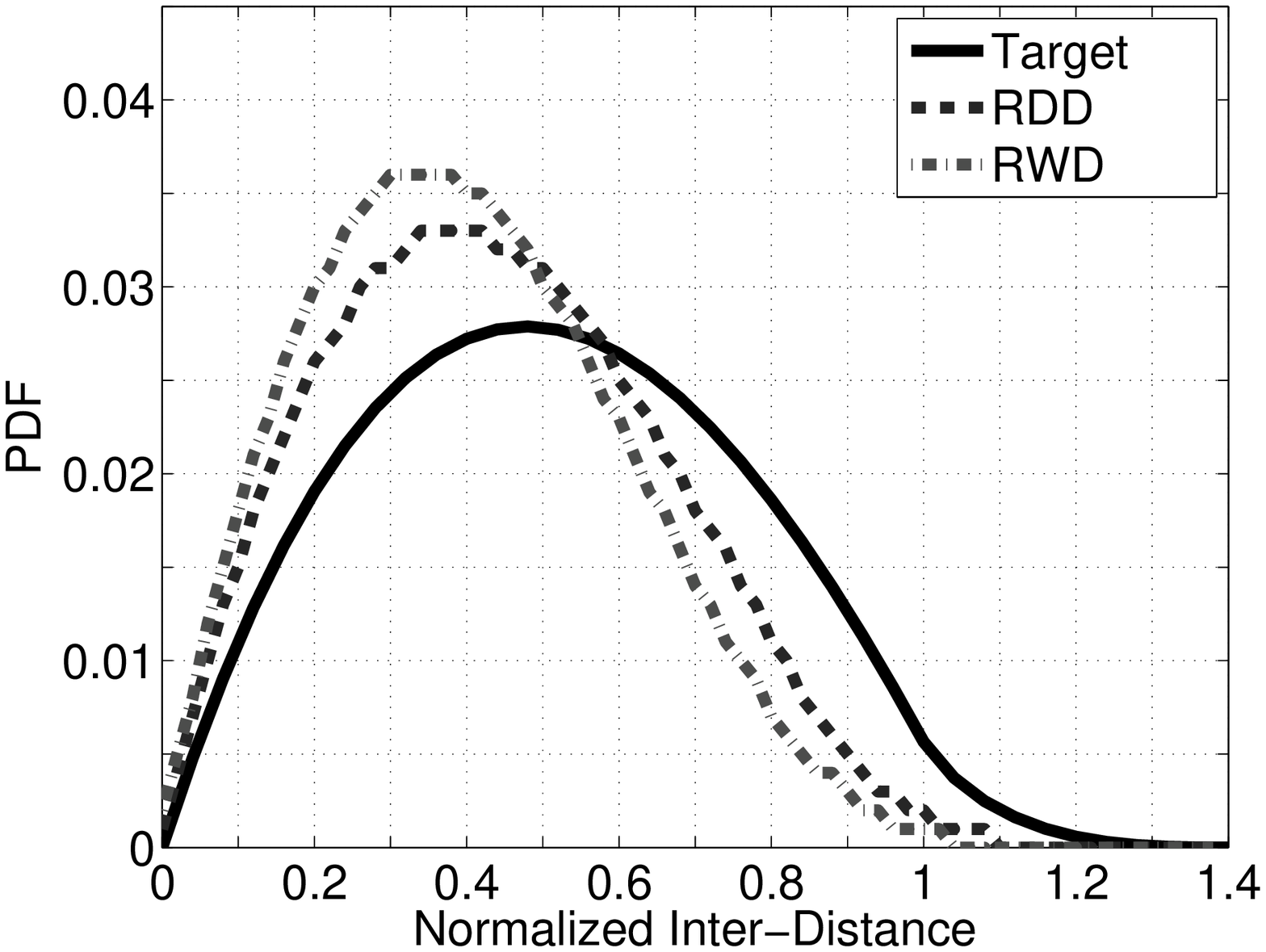}}
      \end{center}
    \end{minipage}
   \end{tabular}
  \caption{PDF of the inter-distance between information copies normalized to $\mathcal{A}$,
for the RWD and RDD dissemination policies in static 
({\em uniform} and \emph{stationary})
and mobile scenarios 
({\em random waypoint}) when $\mathcal{C}(0)=200$ and
    $\tau=10$~s. Here the target distribution is the spatial uniformity.}    
  \label{fig:global-inter-distance-stationary}
 \end{figure*}

\begin{figure*}[htbp]
  \begin{tabular}{cc} 

    \begin{minipage}[t]{0.5\textwidth}
      \begin{center}
        \subfigure[Static scenario]{\includegraphics[scale=0.3]{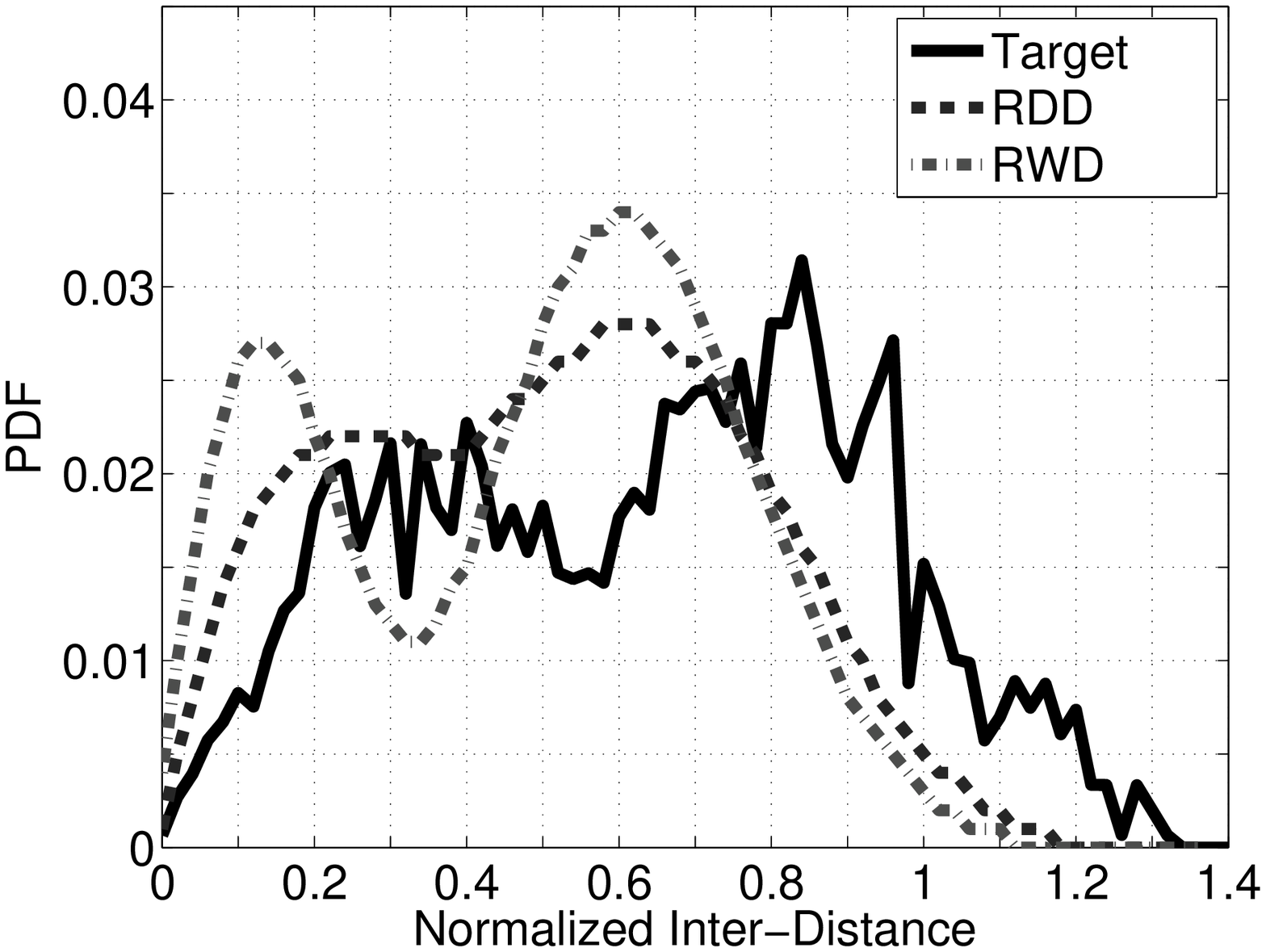}}
        \label{RWD-RDD-static-clustered}
      \end{center}
    \end{minipage}
    
&
    \begin{minipage}[t]{0.5\textwidth}
      \begin{center}
        \subfigure[Mobile scenario]{\includegraphics[scale=0.3]{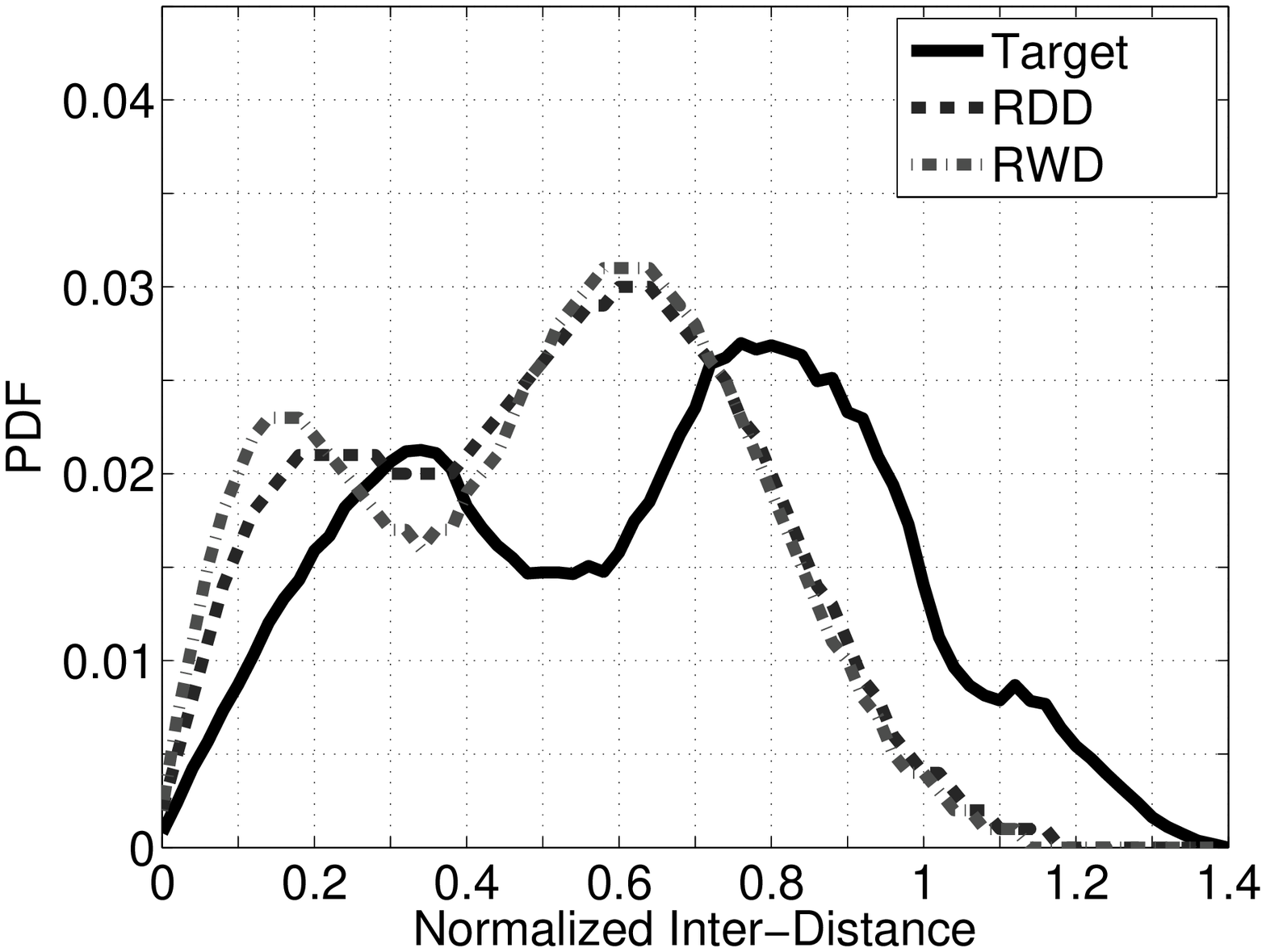}}
        \label{RWD-RDD-mobile-clustered}
      \end{center}
    \end{minipage}

  \end{tabular}
  \caption{PDF of the inter-distance between information copies normalized to 
  $\mathcal{A}$, for the RWD and RDD dissemination policies in static 
  (clustered) and mobile (random trip) 
  scenarios when $\mathcal{C}(0)=200$ and $\tau=10$~s. Here the target distribution is nodal uniformity.}
  \label{fig:global-inter-distance-clustered}
\end{figure*}


Then, we provide a basic performance evaluation of the information query 
process achieved by our application, and focus on the following metrics.

\begin{itemize}
\item {\em Cumulative provider time:}
we evaluate the load balancing properties of the different information 
dissemination strategies by computing the cumulative time $\hat{\tau_i}$ each 
node $i$ spends as an information provider. Given that the cache time $\tau$ is 
deterministic, we can compute $\hat{\tau_i} = \tau \times \mathcal{I}_i$, where 
$\mathcal{I}_i$ accounts for the number of times node $i$ takes up the role of 
information provider during the simulation time.

\item \emph{Served queries at each information provider}: we measure the 
cumulative number of served queries for each information provider $j$. Note that 
this metric is also useful to understand the impact of the hop-based reply 
policy implemented by provider nodes (i.e., the likelihood of replies decreases 
with the increase of hops traversed by the query).

\item \emph{Euclidean distance to access information}: we measure the cumulative 
Euclidean distance from a node to its \emph{closest} information provider, every 
$\tau$ seconds. The distance to access information is the result of 
the spatial distribution of information in the network and can be used to 
measure how ``fair'' our algorithms are toward to each querying node. 

\end{itemize}

\section{The Logistics of Information}
\label{sec:results}

In this section, we look at how the RWD and RDD algorithms we designed can 
achieve the first two objectives outlined in the introduction of the paper: a 
desirable  distribution of the information and a fair distribution of 
information burden across the provider nodes. In the set of results we present, no information drop is 
allowed; indeed, for both the RWD and RDD strategies, a provider that hands the 
information over to another node considers the transfer as successful only if it 
receives an acknowledgment message, otherwise it repeats the procedure by 
selecting a different neighbor. The duplication probability we obtained by 
implementing such an application was negligible (order of $10^{-5}$). Thus, we 
can consider that the overall number of providers does not change  during the 
simulation time (i.e., $\mathcal{C}(t) = \mathcal{C}(0)$); additionally, the 
query rate $\lambda$ is set to a constant value equal to 0.0025 req/s, 
resulting, as discussed in Sec.~\ref{sec:paramters}, in  $\Lambda=4.5$ req/s.


\textbf{Spatial information distribution.} We give an overview of the complete 
set of results for different static node deployments (uniform, stationary and 
clustered) and node mobility models (random waypoint and random trip models), 
and compare the behavior of the RWD and RDD policies.  
In Fig.~\ref{fig:global-inter-distance-stationary} and 
Fig.~\ref{fig:global-inter-distance-clustered}, the target probability density functions (PDF)
correspond to different desired information distributions: in the former the 
target is spatial uniformity, while in the latter 
the objective is nodal uniformity.

Fig.~\ref{fig:global-inter-distance-stationary} shows the PDF of the 
inter-distance between information copies, for both dissemination policies, when 
nodes are deployed according to the stationary distribution and move according to 
the random waypoint model. Similarly, 
Fig.~\ref{fig:global-inter-distance-clustered} shows the PDF of the 
inter-distance between information copies in the static and mobile case when 
nodes are deployed in clusters and move according to the random trip model.
Results are derived for $\mathcal{C}(0) = 200$ and caching time $\tau=10$~s. 
The PDFs are computed from the samples collected over all the simulation time.

In Fig.~\ref{fig:global-inter-distance-stationary} (a),
for the static uniform node deployment, both the RWD and the RDD algorithms
yield an information distribution that perfectly matches the reference 
distribution (overlapping it in the plot).
Under the static stationary scenario, the 
reference PDF is better approximated by the RDD policy than by the RWD policy. 
As for the mobile scenario, not only does node mobility not alter the ``quality'' of the 
approximation achieved by the RDD policy, but it also considerably helps the RWD 
policy in achieving a better information distribution across the geographical 
area, as shown in Fig.~\ref{fig:global-inter-distance-stationary} (b).
Similar observations can be done when we 
consider a clustered scenario, depicted in Fig.~\ref{fig:global-inter-distance-clustered}: 
in the static case, the RDD policy outperforms the RWD policy in better 
approximating the target distribution, while, in the mobile case, the RWD and 
the RDD perform similarly.

Further insights can be gained by observing more closely the behavior of the two 
information dissemination techniques in a simple case: we therefore focus on 
static stationary scenarios and emphasize, using the $\chi^2$ index, the 
differences between the target information distribution and the distribution achieved by the 
RWD and RDD policies. As a target information distribution we take the spatial uniformity. 

\begin{figure}[ptbh]
  \centering
  \includegraphics[scale=0.35]{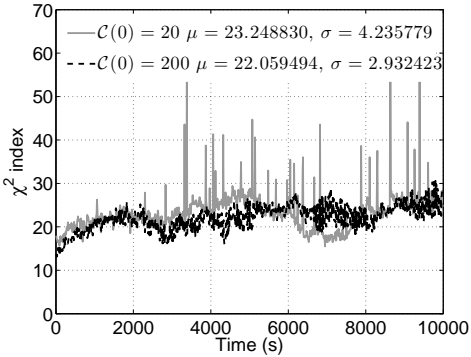}
  \caption{$\chi^2$ index when the RWD model is used in a static 
  stationary scenario: mean ($\mu$) and standard deviation
	         ($\sigma$) with 10 s observation intervals, for $\mathcal{C}(0) = \{20, 200 \}$ and $\tau=10$~s.}
  \label{fig:static_RW_averages}
\end{figure}

\begin{figure}[ptbh]
  \centering
  \includegraphics[scale=0.35]{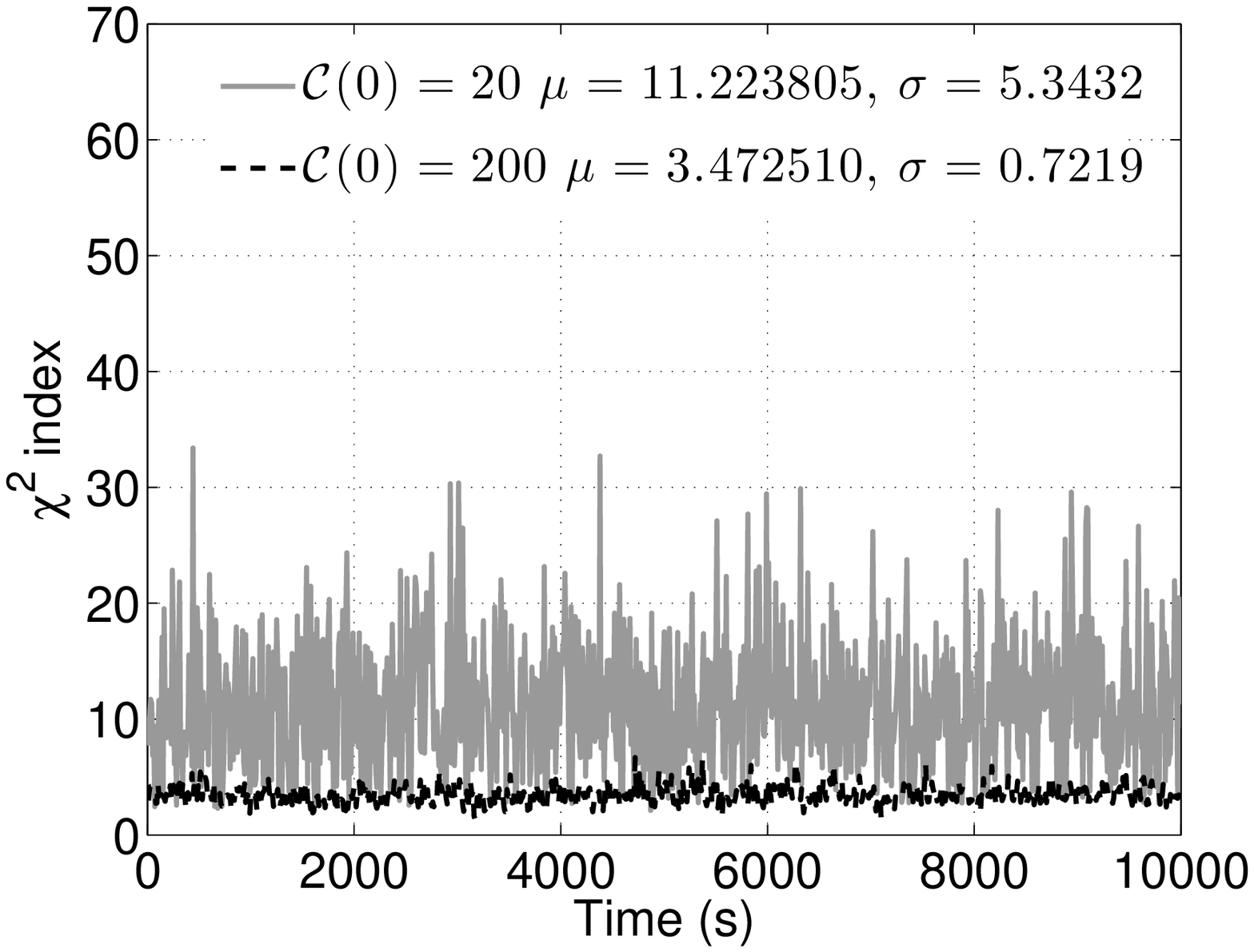}
  \caption{$\chi^2$ index when the RDD model is used in a static 
  stationary: mean ($\mu$) and standard deviation
	         ($\sigma$) with 10 s observation intervals, for $\mathcal{C}(0) = \{20, 200 \}$ and $\tau=10$~s.}
  \label{fig:static_RD_averages}
\end{figure}

The $\chi^2$ index is computed considering the measured and the 
objective probability density function: the smaller the index, the better
the fit.
The evolution of the $\chi^2$ index is plotted over time when the RWD 
and the RDD are applied, respectively, in Fig.~\ref{fig:static_RW_averages} 
and Fig.~\ref{fig:static_RD_averages}. We considered the
number of information copies concurrently moving through the network to sum 
to $\mathcal{C}(0) = 20, 200$ and the caching time to be equal to $\tau=10$~s; also, 
each measured PDF is obtained from  a 10~s observation.
The mean ($\mu$) and standard
deviation ($\sigma$) of the $\chi^2$ index are reported in the legend of 
the figures.

By looking at the plots, we observe that 
varying the number of information providers in the network
dramatically affects the results:
increasing the number of copies from 20 to 200,
the divergence from the theoretical uniform distribution greatly decreases.
Furthermore, 
when compared, the time evolutions of the $\chi^2$ index
of the RWD and RDD algorithms show that the latter better approximates the 
target information distribution.

\begin{figure}[htbp]
  \centering
  \includegraphics[scale=0.35]{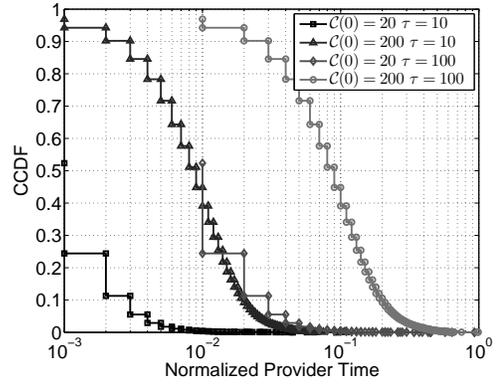}
  \caption{CCDF of the time a node spends in provider mode, normalized to the total simulation time, for the RDD 
policy in a static uniform scenario.}
  \label{fig:provider_time}
\end{figure}



\textbf{Load balancing.} We now turn our attention to the important question of load 
balancing across providers. For brevity, below we present just a 
subset of the results we derived. In particular,  since the RDD manages to provide a better
approximation to the desired information distribution than RWD, 
we only show the performance of the RDD policy. 
Also, we present results only for the static uniform scenario and the mobile network
with random waypoint mobility, 
since a similar performance is achieved under clustered network 
topologies. 

In Fig.~\ref{fig:provider_time} we plot the complementary distribution function 
(CCDF) of the cumulative time a node is serving as an information provider 
(i.e., the provider time) over the whole duration of our experiments, that is, we 
normalize the provider time to the simulation time. The results are presented 
for  the RDD policy in a static uniform scenario, for different values of the 
caching time $\tau$ and when the initial number of information providers sums to 
$\mathcal{C}(0)=20$ and to $\mathcal{C}(0)=200$  (which correspond, respectively, to 
1\% and 10\% of the total number of nodes).

Looking at the figure, we observe that when we increase $\mathcal{C}(0)$ 
from 20 to 200, the load is spread more uniformly 
across the nodes since there is an increased opportunity for being (randomly) 
selected as information provider. The effect of an increased caching time 
$\tau$ from 10 s to 100 s, is, instead, a translation of the CCDF to higher 
values, without affecting the load distribution.

We now look deeper at the impact of different scenarios and simulation 
parameters on the effective load that an information provider supports in terms 
of number of served queries. Note that the number of served queries is not equal 
to the number of queries a provider receives because of the reply behavior 
described in Section \ref{sec:set-up}.

\begin{figure}[htbp]
\begin{minipage}[t]{0.4\textwidth}
      \begin{center}  
\subfigure[$\mathcal{C}(0) = 20$]{\includegraphics[scale=0.35]{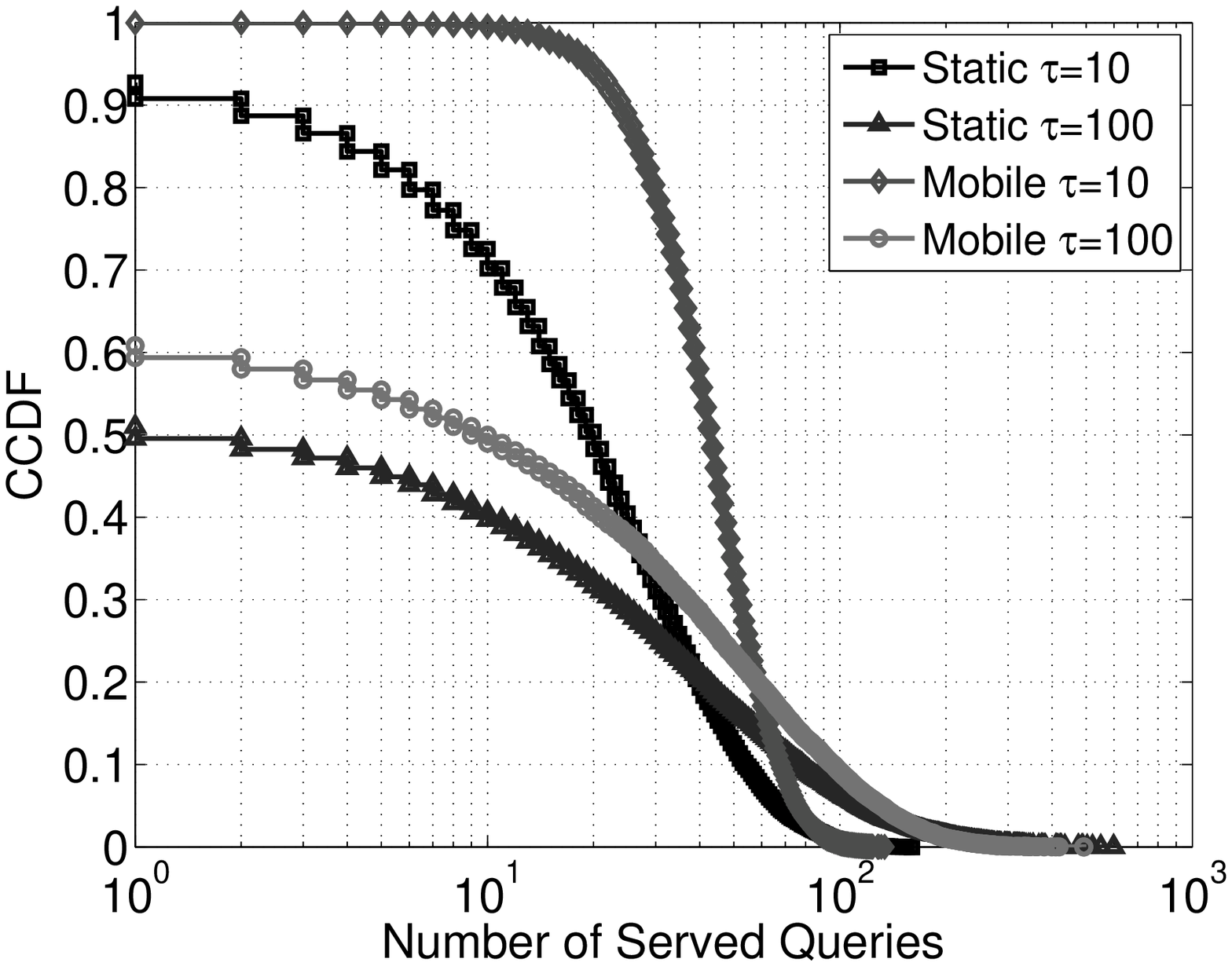}}
        \label{fig:served_queries_20}
      \end{center}
    \end{minipage}

    \begin{minipage}[t]{0.4\textwidth}
      \begin{center}
 \subfigure[$\mathcal{C}(0) = 200$]{\includegraphics[scale=0.35]{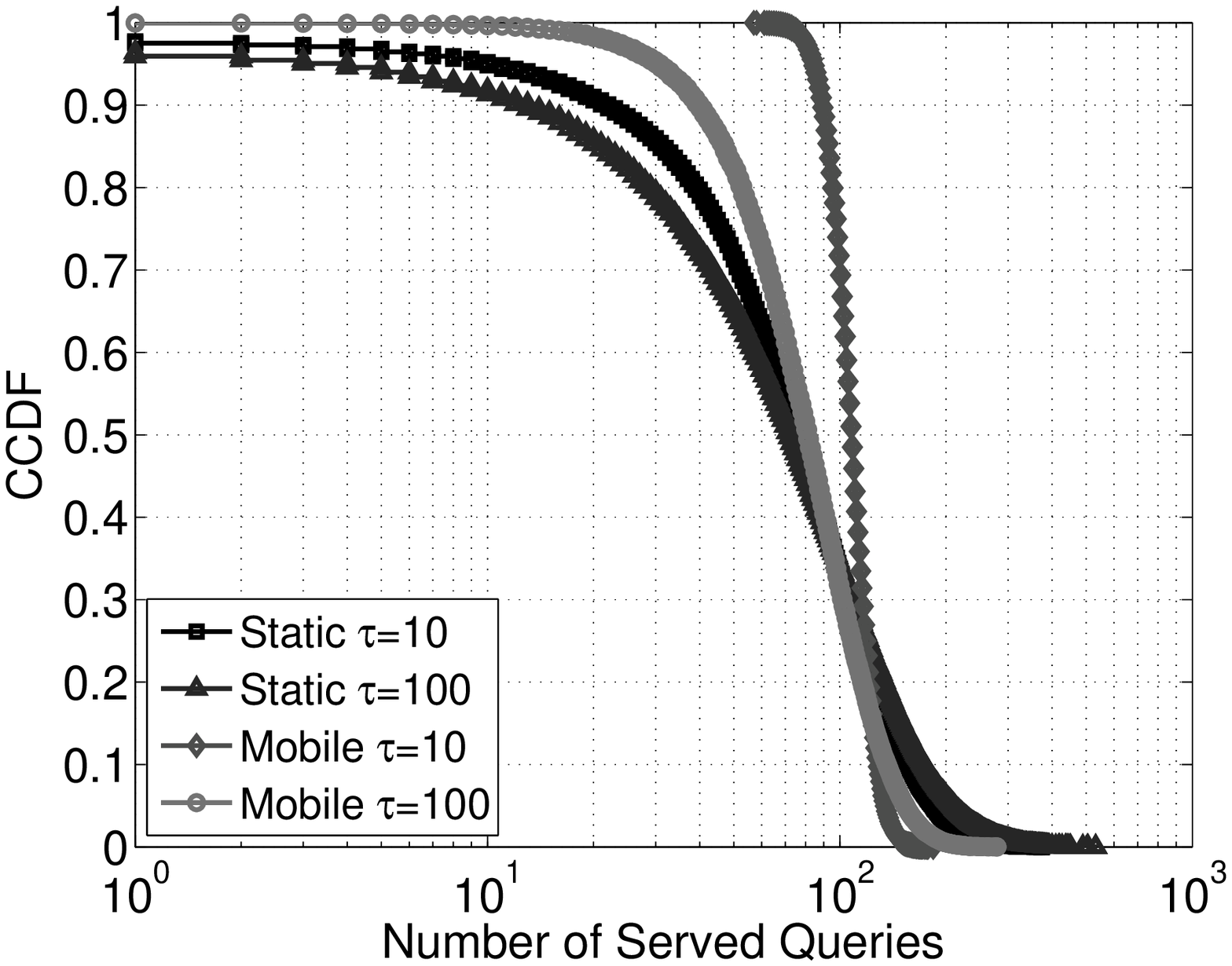}}
        \label{fig:served_queries_200}
      \end{center}
    \end{minipage}

\caption{CCDF of the total number of queries served by information providers for the RDD policy.}
  \label{fig:served_queries}
\end{figure}

Figs.~\ref{fig:served_queries} a) and b) present the CCDF of the number of 
queries served by the provider nodes, respectively,  when $\mathcal{C}(0)=20$ 
and $\mathcal{C}(0)=200$. Both the static uniform scenario and the mobile scenario
with random waypoint mobility are considered. Looking at the plots, we note that an 
increased number of initial providers is effective in spreading the query load 
more evenly, especially in the static case. In the case of the static topology, 
when $\mathcal{C}(0)=20$, roughly 50\% of providers never get a chance to 
satisfy a user request, whereas with $\mathcal{C}(0)=200$, about 60\% of 
providers are serving a number of queries comprised in the interval $[70,150]$. 
The combined effect of node mobility and an increased 
number of initial providers is striking: Fig.~\ref{fig:served_queries} b) 
indicates that approximately 95\% of providers serve roughly the same amount of 
queries. Thus, node mobility, that at a first sight could be considered harmful 
to information distribution mechanisms, turns out to be a good ally in terms of 
load balancing.

\begin{figure}[htbp]
  \centering
  \includegraphics[scale=0.35]{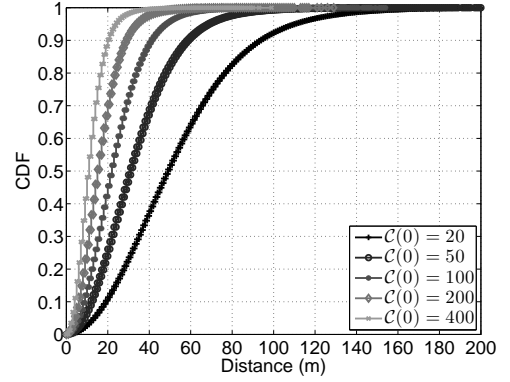}
  \caption{CDF of the Euclidean distance to closest information replica, for the RDD policy in a mobile
    scenario with random waypoint mobility.}
  \label{fig:distance_to_info}
\end{figure}

\textbf{Clients' perspective.} Lastly, we take the perspective of users issuing 
queries to access information held by providers. In 
Fig.~\ref{fig:distance_to_info} we plot the cumulative distribution function 
(CDF) of the Euclidean distance from a querying node to the closest provider, 
for the mobile scenario with random waypoint mobility  and $\tau=10$ s.
More specifically, we study the impact of an 
increasing number of initial providers $\mathcal{C}(0)$, when we let this 
simulation parameter grow from 20 to 400 providers (i.e., from 1\% to 20\% of 
the total number of nodes). Both the mean distance to access information, 
ranging  from 50 m to 15 m, and the variance of the CDF, shrink considerably 
when increasing the number of initial providers. Given that the node radio range 
is set to 20 m, the implications of this result are the following: when a 
sufficient number of initial providers is injected into the network (i.e., 
200--400), nodes may access the information within one hop, whereas an 
insufficient number of initial information copies (i.e., 20--100) may constrain 
a node to use routing to access information that is more than one hop away.

\textbf{Summary.} The evaluation we carried out showed that, despite their simplicity and low overhead,
the proposed algorithms for information dissemination (RWD and RDD) achieve 
the first two objectives defined in this work.
Indeed, as long as enough providers are injected into the network (namely, 
10\% of the total number of nodes), we have that:
\begin{itemize} 
\item[{\em (i)}] under static scenarios, the information distribution yielded by RWD and RDD well
approximates the uniform distribution; in particular the match is excellent with uniform node
deployment for both strategies, while RDD outperforms RWD in stationary and clustered
topologies;
\item[{\em (ii)}] under mobile scenarios,
the approximation is still good for both cases of random waypoint and random trip mobility 
and, interestingly, mobility significantly
improved the RWD performance;
\item[{\em (iii)}] in terms of load balancing, both dissemination strategies
evenly distribute the service load across the provider nodes; again, mobility
has a beneficial effect.
\end{itemize}

\section{Information replication and drop}
\label{sec:replication-drop}
In this section, we factor in a new source of dynamic behavior, i.e.,  
a variable information demand. We thus address the following question:
provided the information demand is uniformly distributed among the nodes,
how can we enhance the basic information dissemination schemes to cope with a
time-varying query rate $\lambda(t)$ and, thus, $\Lambda(t)$?

We propose the following simple modification that applies to both dissemination
policies illustrated in Sec.~\ref{sec:info-distribution}. Whenever the caching
time of an information provider expires, the node decides whether the
information should be \emph{dropped}, \emph{replicated} or if the legacy policy
(RWD or RDD) should be applied. 
In this work we study a simple heuristic for an information provider to take
replication or dropping decisions.

During the bootstrap of the network, the content manager selects an initial 
number of information providers $\mathcal{C}(0)$ and a reference query 
load $\mu^{ref}_{j}$ that each provider is expected to support. Assuming 
$\mu^{ref}_{j} =  \mu^{ref}~\forall j\in\mathcal{C}(0)$, the relationship 
between $\mathcal{C}(0)$ and $\mu^{ref}$ can be expressed as:  

\begin{equation}
 \mathcal{C}(0) \mu^{ref} = (N-\mathcal{C}(0)) \lambda(0) = \Lambda(0)
 \label{eq_mu}
\end{equation}

where $\lambda(0)$ is the initial \emph{per-node} query rate, and
$\Lambda(0)$ is the cumulative query rate, as predicted by the 
content manager. The choice of $\mathcal{C}(0)$ is less straightforward, 
but, as shown in the previous section, a sufficiently high number of 
copies ought to be selected (e.g., 10\% of the total number of nodes). 

During the caching period, provider $j$ keeps track of the number of 
served queries, $\mu_j$. When the caching time expires, $\mu_j$ is compared to
$\mu^{ref}$ and the following actions are taken.

\begin{itemize} 

\item {\em Replication}: if the measured load is higher than the
reference load plus a tolerance factor
$\epsilon$, the information will be replicated and two (as opposed to one) 
new providers are chosen. 

\item {\em Drop}: if the measured load is lower than the
reference load minus a tolerance factor $\epsilon$, 
the information will be dropped and no new provider is selected.

\item {\em Handover}: if the absolute value of the difference 
between measured load and reference load is less than $\epsilon$, the 
legacy behavior is adopted, i.e., the information is handed over to 
another node selected according to the RWD or RDD algorithm.

\end{itemize} 

Depending on the current query rate, we would like the number of providers to 
converge to the ideal value which would approximately maintain the constant 
reference load derived in (\ref{eq_mu}). The number of information
providers should converge to:
\begin{equation}
 \mathcal{C}_{\lambda}(t)  = N \frac{\lambda(t)}{\lambda(t)+\mu^{ref}} 
 \label{eq_C}
\end{equation}

Below, we evaluate the RWD and RDD algorithms enhanced with
the replication/drop heuristic just outlined.

\subsection{Results}
\label{sec:results_replication}
Here we focus on a static network composed by nodes deployed according 
to the stationary distribution defined in Sec.~\ref{sec:set-up}, in which 
the only source of dynamics is due to a time-varying query rate $\Lambda(t)$. 

In the following, we set the simulation parameters
to $N=2000$, $\tau = 100$ s, $\mathcal{C}(0) = 200$, $\Lambda(0) = 4.5$ req/s 
while the simulation time is 20000 seconds.
We simulate a time-varying \emph{per node} query rate by dividing the simulation 
time in four phases: 
\begin{enumerate}
\item  $\lambda(t)$ increases by $0.6$ \% every second 
when $t \in [1,2500)$ s; 
\item  $\lambda = 1/100$ req/s when $t \in [2500,10000)$ s; 
\item  $\lambda(t)$ decreases by $0.3$ \% every 
second when $t \in [10000,12500)$ s; 
\item  $\lambda = 1/200$ req/s  when $t \in [12500,20000]$ s.
\end{enumerate}

Given $\mathcal{C}(0)=200$ providers, using (\ref{eq_mu}) we set 
$\mu^{ref} = 0.0025$ req/s as the reference query load that each information
provider is requested to satisfy.

Equation~(\ref{eq_C}) helps in predicting the ideal number of providers that should emerge 
from the replication/drop process: in phase 2), the ideal number of providers
should add up to $\mathcal{C}_{\lambda}(t) \approx 616$, whereas in phase 4)
$\mathcal{C}_{\lambda}(t) \approx 364$ providers.  

\begin{figure}[!t]
  \centering
  \includegraphics[scale=0.35]{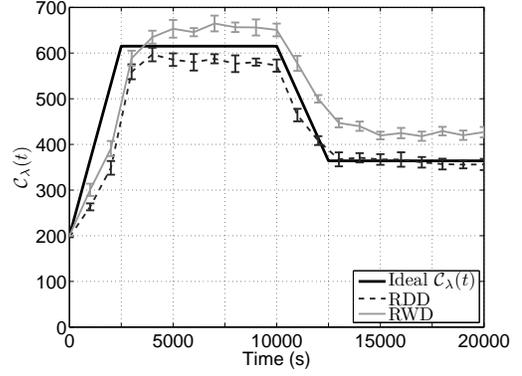}
  \caption{Time evolution of the number of information providers $\mathcal{C}(t)$ in the static 
stationary case, for a time-varying $\lambda(t)$.}
  \label{fig:rep_static}
\end{figure}

Fig.~\ref{fig:rep_static} refers to the static stationary scenario and
presents the time evolution of the number of providers in the network $\mathcal{C}(t)$ 
collected at 1,000 s intervals,
each point averaged over 10 simulation runs.
A 95\% confidence interval is also shown. 
We observe that, as the query rate increases, the replication process takes 
place and the number of information providers increases, following the 
ideal number of providers reported in the figure.
During steady state, providers shift to the legacy handover behavior, 
whereas when the query rate starts its descent, the dropping process 
is selected and the number of providers decreases, until a new steady state 
regime is reached.

When the query rate settles to a steady state (phase 2) and 4)), 
the replication/drop heuristic behaves differently if applied to the RWD
or the RDD mechanisms. The RWD policy consistently overestimates by roughly
10 \% the number of ideal providers; instead, the RDD policy is very
accurate. 
The discrepancy between the RDD and RWD policies can be attributed to the 
poorer ability of the latter mechanism to approximate a uniform distribution 
of the information, as pinpointed in Sec.~\ref{sec:results}. 
The consequence is that the replication/drop process will be 
altered due to an uneven query load at providers, hence the better 
performance of the RDD mechanism.

We also performed a series of experiments for a scenario in which nodes 
move according to the random waypoint mobility model. The behavior of
the replication/drop heuristic is comparable to that of a static scenario:
however, similarly to our findings in Sec.~\ref{sec:results}, we observe that
mobility helps in spreading the query load among providers, hence the
number of providers that emerge in the replication/drop process is
a better approximation of the ideal value.


\section{Conclusions}

We considered a peer-to-peer wireless network, where 
nodes may act as both clients and providers to other network nodes.
In such a cooperative environment, we addressed the problem of achieving a desired 
distribution of information and  
a fair load distribution among the provider nodes.
We designed low-overhead, content-transparent, distributed algorithms
that regulate the information storage at the network nodes and allow
a fair selection of the nodes acting as providers.

Our experimental results indicated that under a variety of scenarios
including static, mobile, and clustered network topologies, our simple
mechanisms were effective in approximating a desired information distribution.
Mobility appeared to be a useful ally, instead of a problematic phenomenon,
since it helped to achieve an even distribution of the load on providers.

We also challenged our dissemination mechanisms with a time-variant 
information demand, and proposed a simple enhancement to achieve a constant,
predefined load imposed at each provider despite this additional source
of dynamics. Experiments showed that a quick convergence to an ideal 
number of providers can be achieved in a simple, distributed way.

\end{document}